\documentclass[%
aip,
jmp,%
amsmath,amssymb,
preprint,
]{revtex4-1}

\usepackage{graphicx}% Include figure files
\usepackage{bm}% bold math
\usepackage{ulem}
\newcommand{\fref}{Fig.~\ref}
\newcommand{\eref}{Eq.~\ref}

\usepackage{color}

\begin{document}

\title{What is the best planar cavity for maximizing coherent exciton-photon coupling}

\author{Zhaorong~Wang} \affiliation{EECS Department, University of Michigan, Ann Arbor, MI 48109, USA}
\author{Rahul~Gogna} \affiliation{Applied Physics Department, University of Michigan, Ann Arbor, MI 48109, USA}
\author{Hui~Deng} \email{dengh@umich.edu}
\affiliation{Physics Department, University of Michigan, Ann Arbor, MI 48109, USA}


\begin{abstract}
We compare alternative planar cavity structures for strong exciton-photon coupling, where the conventional distributed Bragg reflector (DBR) and three unconventional types of cavity mirrors -- air/GaAs DBR, Tamm-plasmon mirror and sub-wavelength grating mirror. We design and optimize the planar cavities built with each type of mirror at one side or both sides for maximum vacuum field strength. We discuss the trade-off between performance and fabrication difficulty for each cavity structure. We show that cavities with sub-wavelength grating mirrors allow simultaneously strongest field and high cavity quality.
The optimization principles and techniques developed in this work will guide the cavity design for research and applications of matter-light coupled semiconductors, especially new material systems that require greater flexibility in the choice of cavity materials and cavity fabrication procedures.
\end{abstract}

%\pacs{42.50.Pq, 42.50.Nn, 42.55.Sa, 42.79.Dj}% PACS, the Physics and Astronomy
%\keywords{microcavity; grating; photonic crystals; polariton}%Use showkeys class option if keyword

\maketitle

%\section{Introduction}
Strong coupling between semiconductor quantum well (QW) excitons and cavity photons leads to new quasi-particles -- microcavity polaritons. Since their discovery \cite{weisbuch_observation_1992}, planar microcavity polaritons have become a fruitful ground for research on fundamental cavity quantum electrodynamics, macroscopic quantum coherence, and novel device applications \cite{Khitrova_Nonlinear_1999, deng_exciton-polariton_2010,sanvitto_road_2016}.
Crucial for polariton research is a cavity with a strong vacuum field strength $E$ at the QW plane and a high quality factor $Q$. A strong vacuum field leads to stronger exciton-photon coupling and thus a larger vacuum Rabi splitting between the polariton modes. 
A high quality factor leads to long lifetime and coherence time of the cavity photon and correspondingly the polariton. They together enable robust polariton modes, thermodynamic formation of quantum phases, and polariton lasers with lower density threshold at higher operating temperatures \cite{deng_exciton-polariton_2010}.

Given the importance of a strong cavity field, there has been a few decades of effort to improve the cavity field strength. Most commonly used in polariton research are planar Fabry-Perot (FP) cavities formed by two distributed Bragg reflectors (DBRs), each DBR consisting of either epitaxially-grown, closely lattice-matched alloys or amorphous dielectric layers. This type of cavity can reach $Q$ of tens of thousands in III-As based systems \cite{Reitzenstein_AlAs_2007,Nelsen_Dissipationless_2013}, but only about a thousand for III-N \cite{christmann_room_2008,das_room_2011} and even lower for other material systems \cite{kena-cohen_room-temperature_2010,liu_strong_2015}. The field penetrates many wavelengths into the DBR, hence the field strength is not optimal \cite{Yamamoto_Microcavity_1991, Norris_Time-resolved_1994}.

To achieve a stronger field, different cavity structures have been developed with reduced effective cavity length, including using a metal mirror\cite{Grossmann_Tuneable_2011} and using Al-oxide \cite{Nelson_Room-temperature_1996, Graham_Exciton_1997, Pratt_Photoluminescence_1999} or air/GaAs DBRs \cite{Grossmann_Tuneable_2011, Gessler_Low_2014} with larger index contrast. These structures showed greater polariton splitting but unfortunately worse Q compared to AlGaAs-based DBRs; polariton lasing have not been possible in these structures. Recently, cavities using a high index-contrast sub-wavelength grating (SWG) \cite{huang_surface-emitting_2007} were demonstrated for polariton lasers \cite{zhang_zero-dimensional_2014}. Not only does SWG allow greater design flexibility \cite{Wang_Dispersion_2015, Zhang_Coupling_2015} and compatibility with unconventional materials, it is only a fraction in thickness compared to typical DBRs, making it promising for shorter effective cavity length. Yet the demonstrated SWG-based polariton cavity was not optimized for strong cavity field.

Here we demonstrate how to optimize these different types of cavity for the strongest field, and compare their polariton splitting, quality factors and practicality for fabrication. In particular, we find SWG cavities may allow simultaneously stronger field and high Q.

The cavity field relevant to the exciton-photon coupling is the vacuum fluctuation field. Its strength $ E $ is normalized to the zero-point photon energy by 
\begin{equation}\label{eq:normalization}
	\int{\epsilon(\bm{r}) E(\bm{r})^2 dV }= \frac{1}{2}\hbar \omega_c ,
\end{equation}
where $ \epsilon $ is the permittivity, and $ \hbar\omega_c $ is the cavity resonance energy. The integral is evaluated in the entire space with a spatial-dependent $ \epsilon $. For a planar cavity, we consider the field being confined in the $ z $-direction but unbounded in the $ x $ and $ y $ directions. The integration is thus evaluated in a quantization volume ($ L_x $, $ L_y $, $ L_z $), where $ L_x $ and $ L_y $ are set to be much larger than the wavelength, and $ L_z $ is set at some cutoff point where the field has decayed significantly ($ \sim $1\%).

The maximum field strength $ E_{max} $ serves as our main figure of merit for cavities designed for strong coupling. Specifically, for a QW placed at the field maximum, the vacuum Rabi splitting $\hbar\Omega$ is directly proportional to $ \bm{d}\cdot \bm{E}_{max} $, for $\bm{d}$ the QW exciton dipole moment. We note that the $ E_{max} $ is closely related to the mode volume commonly used to characterize photonic crystals cavity\cite{Foresi_Photonic-bandgap_1997, Kristensen_Generalized_2012} as well as the effective cavity length $L_c$ for planar cavities\cite{Yamamoto_Microcavity_1991}. We choose $E_{max}$ over $L_c$ because the vacuum Rabi splitting is determined by not only $ L_c $ but also $ \epsilon $ as $ \hbar\Omega \propto (\epsilon L_c)^{-1/2}  $,\cite{Norris_Time-resolved_1994} while $\epsilon$ can vary by an order of magnitude in different cavities. Therefore the maximum vacuum field strength is the most unambiguous quantity to characterize the cavity for strong coupling. Also, we focus on high-Q cavities but do not optimize for Q. This is because for cavities designed to have sufficiently high Q, the experimentally achievable Q depends mainly on practical constrains such as chemical purity and structural integrity of the fabricated structure.

In the following, we optimize the maximum vacuum field strength $ E_{max} $ for different types of planar cavities. Based on \eref{eq:normalization}, the $E_{max}$ is enhanced in cavities with (i) a tightly confined field profile $E(\bm{r})$ and (ii) materials of low refractive indices. A tightly confined field profile effectively reduces the range of integration, enhancing the field maximum. For a planar cavity, this means using the shortest possible cavity length (half wavelength) and using mirrors with shorter field penetration length.
Equally important, lower refractive indices lead to higher $E(\bm{r})$, especially in the cavity layer where most of the field resides. Vacuum or air has the lowest refractive index. Although not applicable to crystalline QWs that needs mechanical support and surface protection, air-cavities can greatly enhance field strength for QWs formed by two-dimensional (2D) van der Waals materials.

We use transfer matrix method for calculations of DBR-based cavities, and rigorous coupled-wave analysis (RCWA) for SWG-based cavities. We obtain the cavity resonance from the reflection spectrum of the cavity, using the real-valued bulk permittivity $\epsilon_\infty$ for the QW layer. Then we compute complex field distribution at the resonance and normalize the field using \eref{eq:normalization}. To include the exciton-photon coupling, we use a linear dispersion for the QW layer\cite{Zhu_Vacuum_1990} modeled by a Lorentz oscillator\cite{Houdre_Room-temperature_1994}:
\begin{equation}
\label{eq:LorentzDispersion}
\epsilon(e) = n^2(e) = \epsilon_\infty + \frac{fq^2\hbar^2}{m\epsilon_0L_z}\frac{1}{e_0^2-e^2 - i\gamma e},
\end{equation}
where $ f $ is the oscillator strength per unit area, $ q $ and $ m $ are the charge and mass of the electron respectively, $ L_z $ is the QW thickness, $ e_0 $ is the exciton energy, and $ \gamma $ is the exciton linewidth.
For a fair comparison of the different types of cavities, we focus on 2D half-wavelength cavities with a single QW placed at the field maximum.\footnote{In all cavities, multiple QWs can be used to enhance the polariton splitting.}
We use III-As systems as an example, because polariton cavities of highest qualities are all based on III-As systems. The cavity is made of $\rm Al_x GaAs$ alloys where the different Al content gives different refraction index, with AlAs the lowest ($n=3.02$), GaAs the highest ($n=3.68$), and $\rm Al_x GaAs$ a linear interpolation ($n= 3.68-0.66 x $). For the QW, we consider either a 12~nm GaAs/AlAs QW or a 7~nm InGaAs/GaAs QW. For simplicity and ease of comparison, we assume both to have the same exciton energy of 1.550~eV (800~nm)\cite{zhang_zero-dimensional_2014}, oscillator strength of $ f=6\times10^{-4} \AA^{-2} $ ~\cite{Andreani_Accurate_1990} and $\gamma=0.8$~meV.

\begin{figure}[ht]
	%\flushright % declaration
	\includegraphics[width=0.45\textwidth]{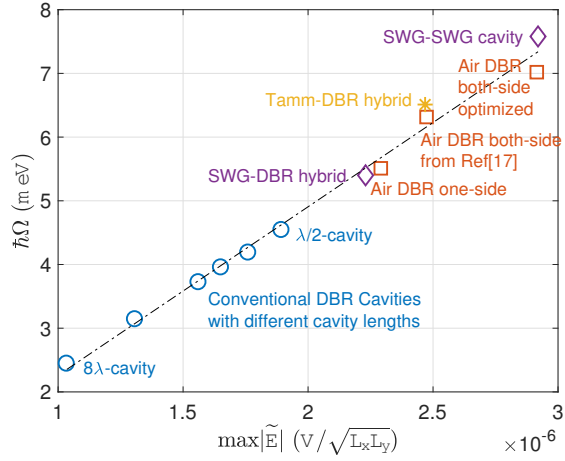}
	\caption{Calculated vacuum Rabi splitting $ \hbar\Omega $ and maximum vacuum field strength $ max |\widetilde{E}| $ for various optimized cavities. All cavities are based on III-As material system. Single QW is embedded at the field maximum. The vacuum field strength is calculated at the cavity resonance energy with QW dispersion turned off, while the Rabi splitting is measured from the two reflection dips when the QW dispersion is turned on. The dash-dotted line is a linear fit through all points.
	}
	\label{fig:Rabi_vs_field_various_cavities}
\end{figure}	
As a benchmark, we first describe the performance of the most widely used, monolithic DBR-DBR cavities. DBRs are typically made of multiple pairs of high- and low-index layers, all with an optical path length (OPL) of $ \lambda/4 $, for $\lambda$ the cavity resonance wavelength. High reflectance is achieved by maximal constructive interference among multi-reflections from the layer interfaces.
Light decays in a distance inversely proportional to the refraction index contrast of the DBR pair. So a high index contrast is preferred for the DBR pair. For a monolithic III-As cavity with a GaAs QW, we use $ \rm AlAs$ ($n= 3.02$) for the low-index layer and $ \rm Al_{0.15}Ga_{0.85}As$ ($n=3.58$) for the high index layer, where $15\%$ AlAs alloy is included instead of pure GaAs to avoid absorption at the QW exciton resonance. We consider a bottom DBR with 20 pairs on GaAs substrate, and a top DBR with 16.5 pairs, matching the reflectance of the bottom DBR. An experimental cavity $Q$ of a few thousands can be readily achieved with this structure. More DBR pairs will lead to a higher $Q$, but will have negligible effect on the field confinement or the vacuum Rabi splitting.

\begin{figure}[t]
	%\flushright % declaration
	\includegraphics[width=0.235\textwidth]{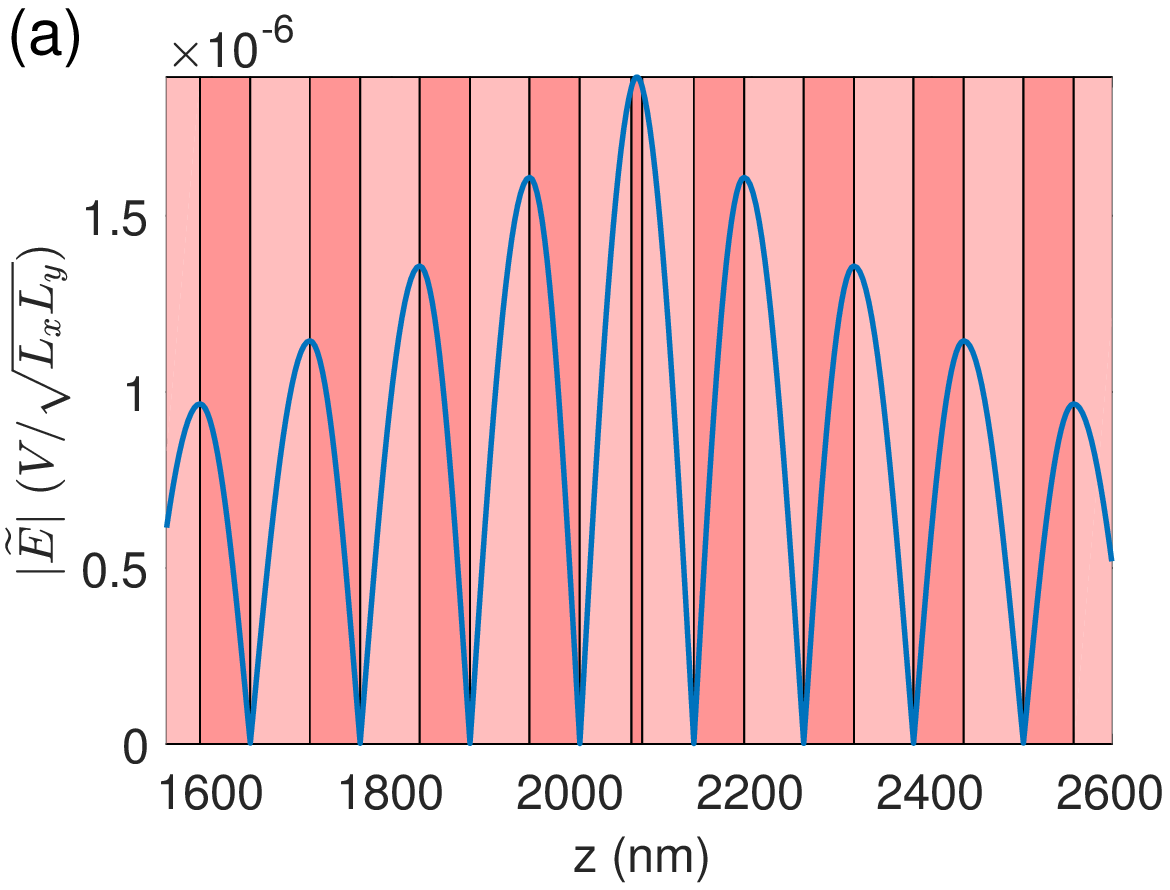}
	\includegraphics[width=0.235\textwidth]{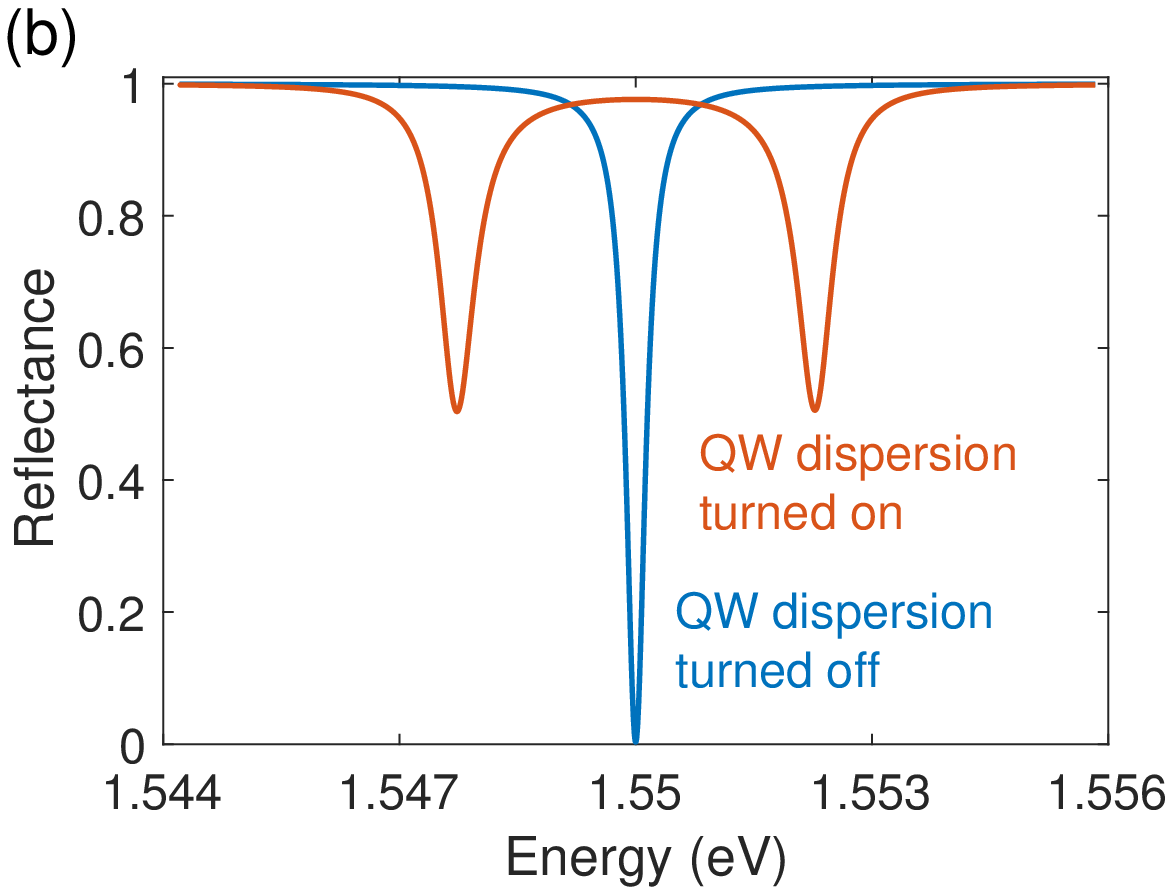}
	\includegraphics[width=0.235\textwidth]{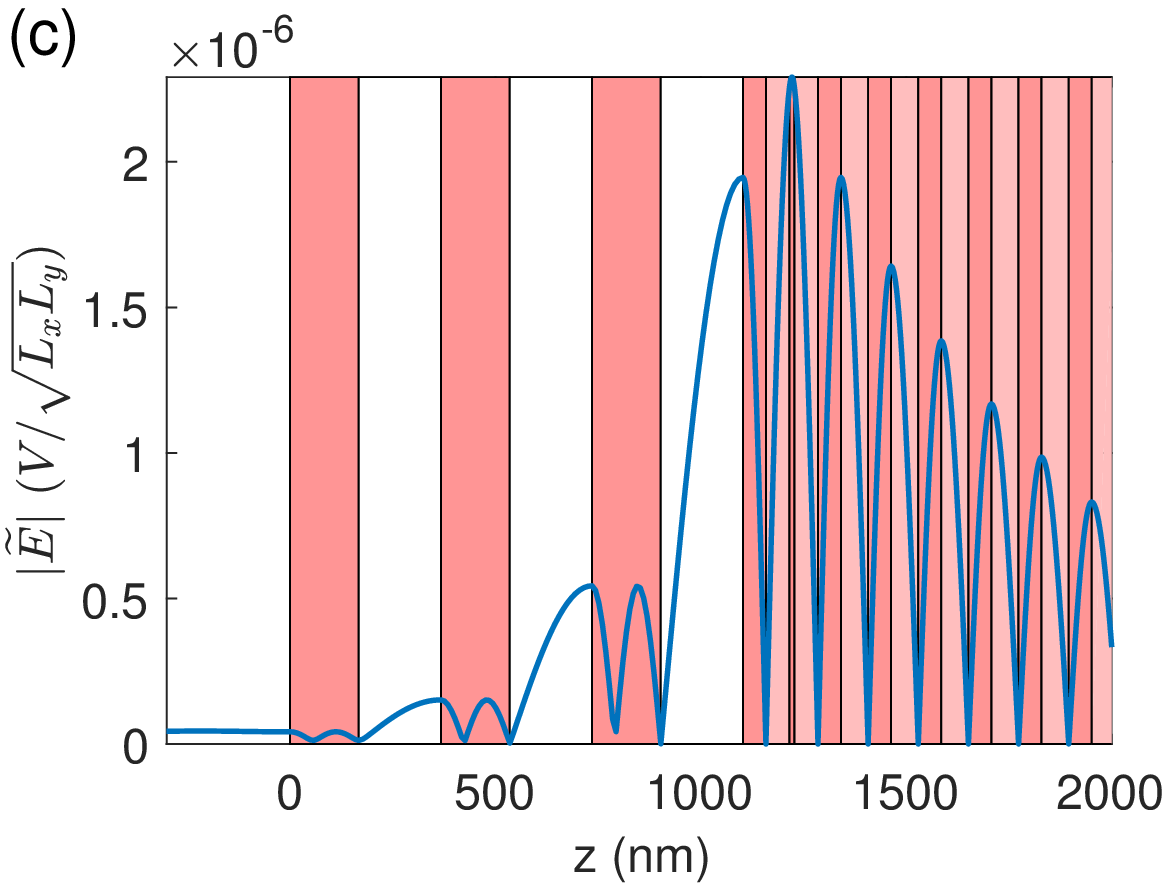}
	\includegraphics[width=0.235\textwidth]{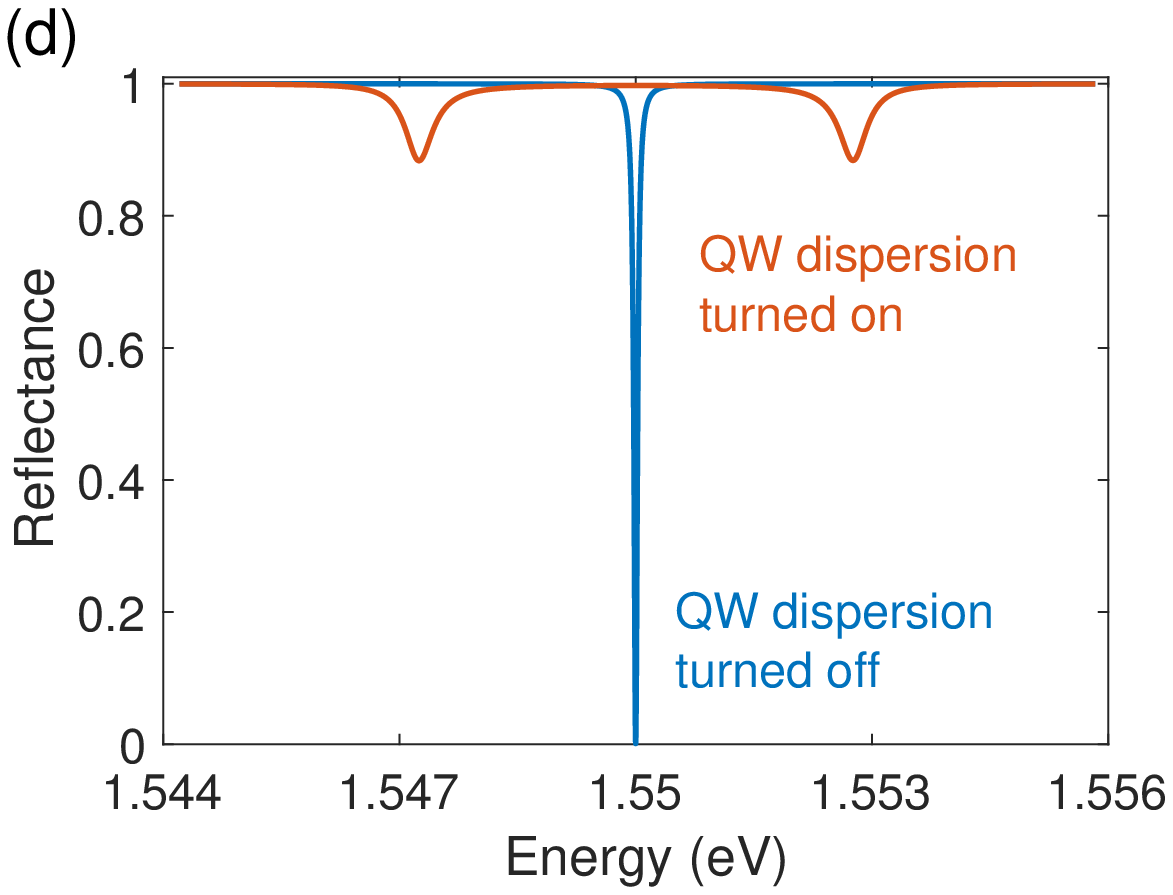}
	\includegraphics[width=0.235\textwidth]{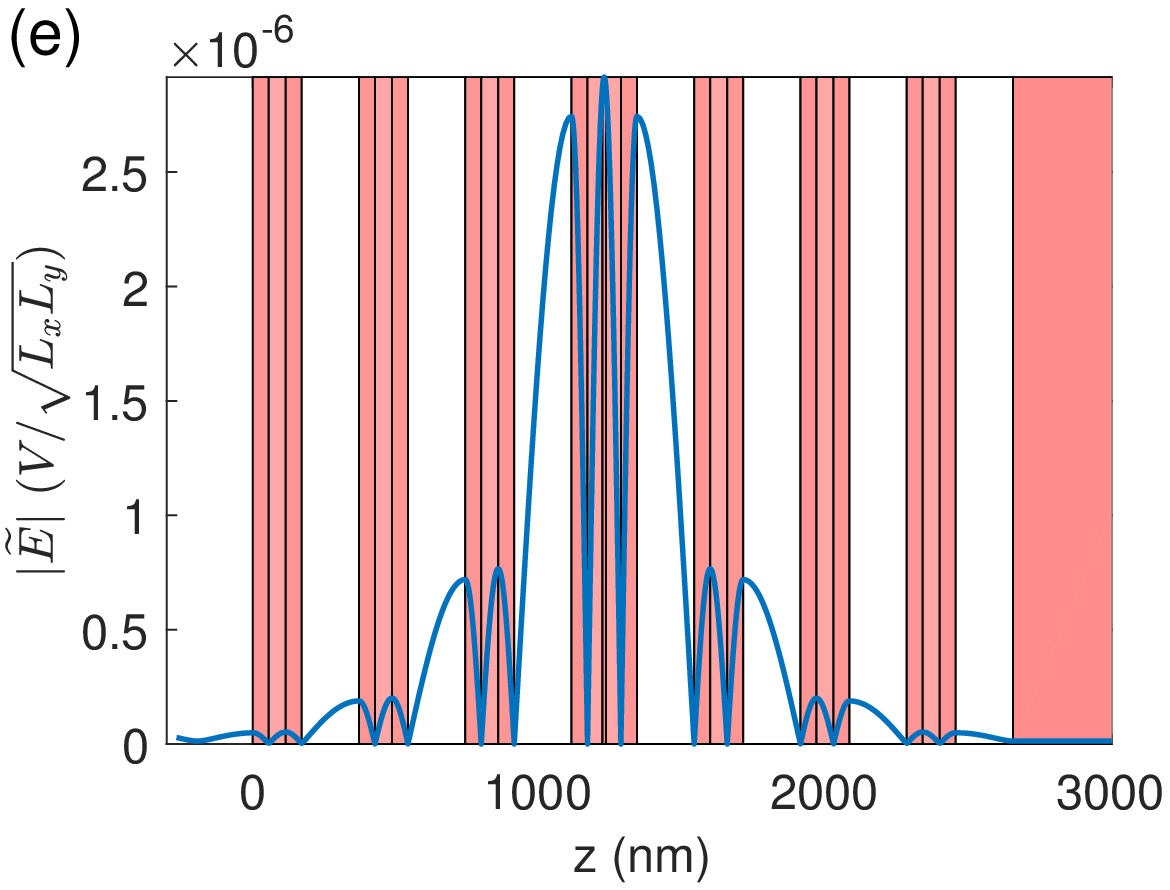}
	\includegraphics[width=0.235\textwidth]{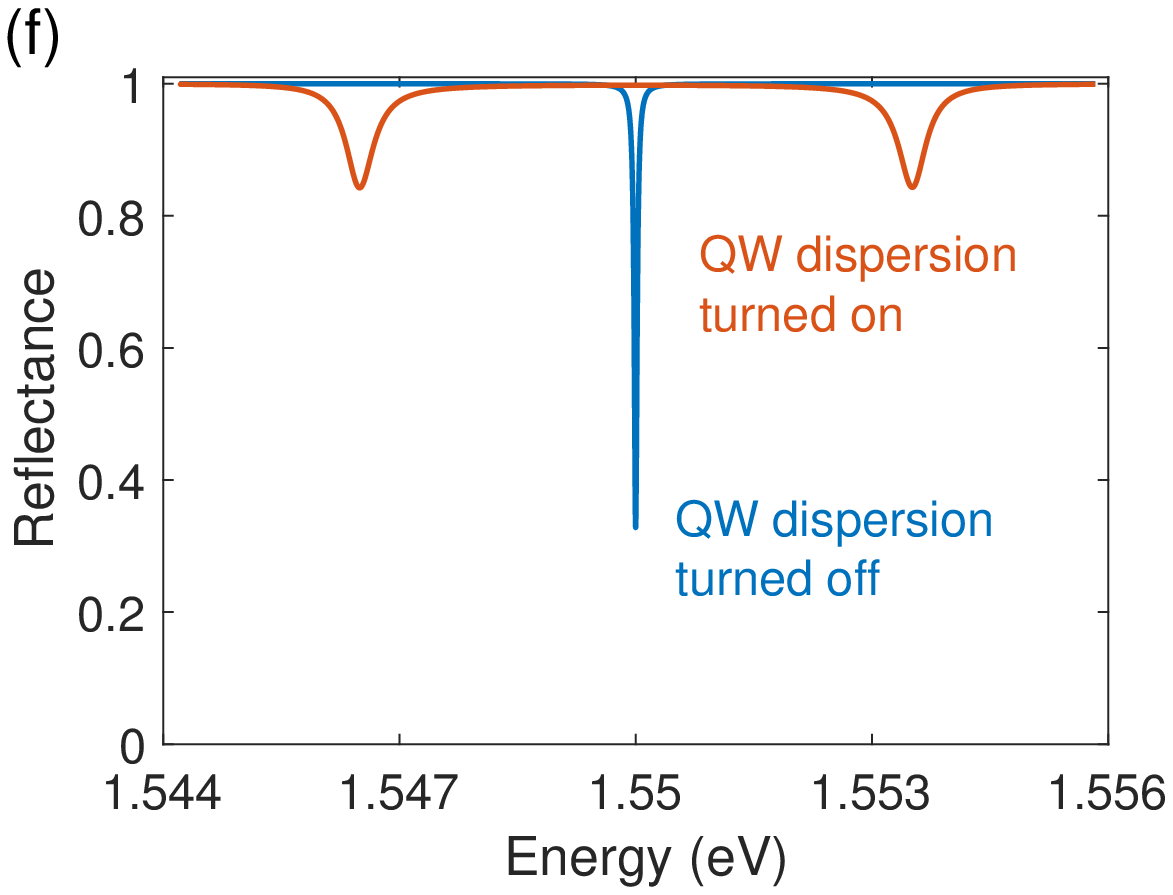}
	\caption{Vacuum field profile and Rabi splitting for conventional DBR cavity (a-b) and air DBR cavity (c-f). The color stripes represent different materials-- red is $ \rm Al_{0.15}Ga_{0.85}As$, lighter red is the lower index $ \rm AlAs$, white region is air, and the thin stripe at the field maximum is a GaAs QW. All three structures sit on GaAs substrate, which is not shown in (a) and (c). The vacuum field strength has a unit of Volt/(normalization length in meters).
	}
	\label{fig:air_DBR_cavity}
\end{figure}

To show the enhancement of the cavity vacuum field strength by reducing cavity length, we vary the cavity layer OPL from $ \lambda/2 $ to $8\lambda$. As shown in \fref{fig:Rabi_vs_field_various_cavities}, $E_{max}$ increases with decreasing cavity length, and the polariton splitting scales linearly with $ E_{max} $. This confirms the $E_{max}$ as an appropriate figure of merit for optimizing exciton-photon coupling. The field distribution of the best-performing $ \lambda/2 $ cavity is shown in \fref{fig:air_DBR_cavity}(a). The maximum field strength is $E_{max}= 1.89\times 10^{-6} \rm V/\sqrt{L_x L_y} $ at the QW, but the field extends many wavelengths into the DBRs due to the relatively small index contrast ($<1.2:1$) of the DBR layers. The reflectance spectrum shows a polariton splitting of 4.54~meV (\fref{fig:air_DBR_cavity}(b)). This conventional cavity can be fabricated by mature epitaxial growth technology as a monolithic crystalline structure with few impurities or defects. However, the requirement of lattice-matched materials, with resultantly small index contrast, leads to limited field confinement.

For the second type of cavity, we increase the index contrast of the DBR layer by replacing the AlAs layer by air or vacuum ($n=1$), which we call an air-DBR. It can be created via selective wet etching \cite{Grossmann_Tuneable_2011,Gessler_Low_2014}. It represents the the highest possible refraction index contrast for a DBR. The GaAs QW still needs to be embedded inside an AlAs cavity (as opposed to an air cavity) to avoid detrimental surface effects. Each material layer -- both the cavity layer and the $\rm Al_{0.15}GaAs$ layers in the DBRs, need to be sufficiently thick for mechanical stability. The experimentally realized structure used a minimum OPL of $ 3\lambda/4 $ for each suspended layer \cite{Gessler_Low_2014}, which we also adopt. Based on this, \fref{fig:air_DBR_cavity}(c) shows the optimal structure we obtained. The top mirror consists of 3 pairs of $ \rm Al_{0.15}Ga_{0.85}As$/air layers. The air layer has $ \lambda/4 $ OPL, the $ \rm Al_{0.15}Ga_{0.85}As$ layer has $ 3\lambda/4 $ OPL ($ \sim 168$~nm). The bottom DBR is the same as in the DBR-DBR cavity (\fref{fig:air_DBR_cavity}(a)). The maximum field strength is increased to $E_{max}= 2.29\times 10^{-6} \rm V/\sqrt{L_x L_y} $. The corresponding vacuum Rabi splitting is $\hbar\Omega = 5.51$~meV, 21\% larger than the conventional DBR cavity.

Further increase of the field confinement can be obtained by replacing the bottom DBR also by an air-DBR, as shown in \fref{fig:air_DBR_cavity}(e). 
An experimentally fabricated air-DBR cavity \cite{Gessler_Low_2014} used a 7~nm $\rm In_{0.13}GaAs $ QW embedded in a GaAs high-index $ \lambda$-cavity. For this structure, we calculated $E_{max}=2.47\times 10^{-6} \rm V/\sqrt{L_x L_y} $ and $\hbar\Omega =6.32$~meV (\fref{fig:Rabi_vs_field_various_cavities}), which is 39\% improvement over the conventional DBR cavity.

The structure can be further optimized by using a cavity layer of low refraction index. However, an AlAs-cavity is incompatible with the selective wet-etching required to create the bottom air-DBR; instead a Ga-rich layer is needed. Hence, for the cavity, we use a $ \rm Al_{0.4}Ga_{0.6}As$ ($ n_r = 3.33 $) layer of $\lambda/2$ OPL sandwiched between two $ \rm Al_{0.15}Ga_{0.85}As$ ($ n_r = 3.58 $) layers of $ \lambda/4 $ OPL, for a total OPL of $ \lambda $. Similarly the $ 3/4\lambda $-OPL high-index DBR layer can also be replaced by such sandwich structure to improve the index contrast. The wet etching selectivity of AlAs over $ \rm Al_{0.4}Ga_{0.6}As$ has been shown to be $10^7.$\cite{Yablonovitch_Extreme_1987} A 12~nm GaAs QW is placed at the center of the cavity.
We obtain $E_{max}= 2.91\times 10^{-6} \rm V/\sqrt{L_x L_y} $ and $\hbar\Omega =7.02$~meV, 55\% larger than conventional DBR cavity. This structure represents the best Rabi splitting in realistic DBR-based III-As cavities.

\begin{figure}[ht]
	%\flushright % declaration
	\includegraphics[width=0.235\textwidth]{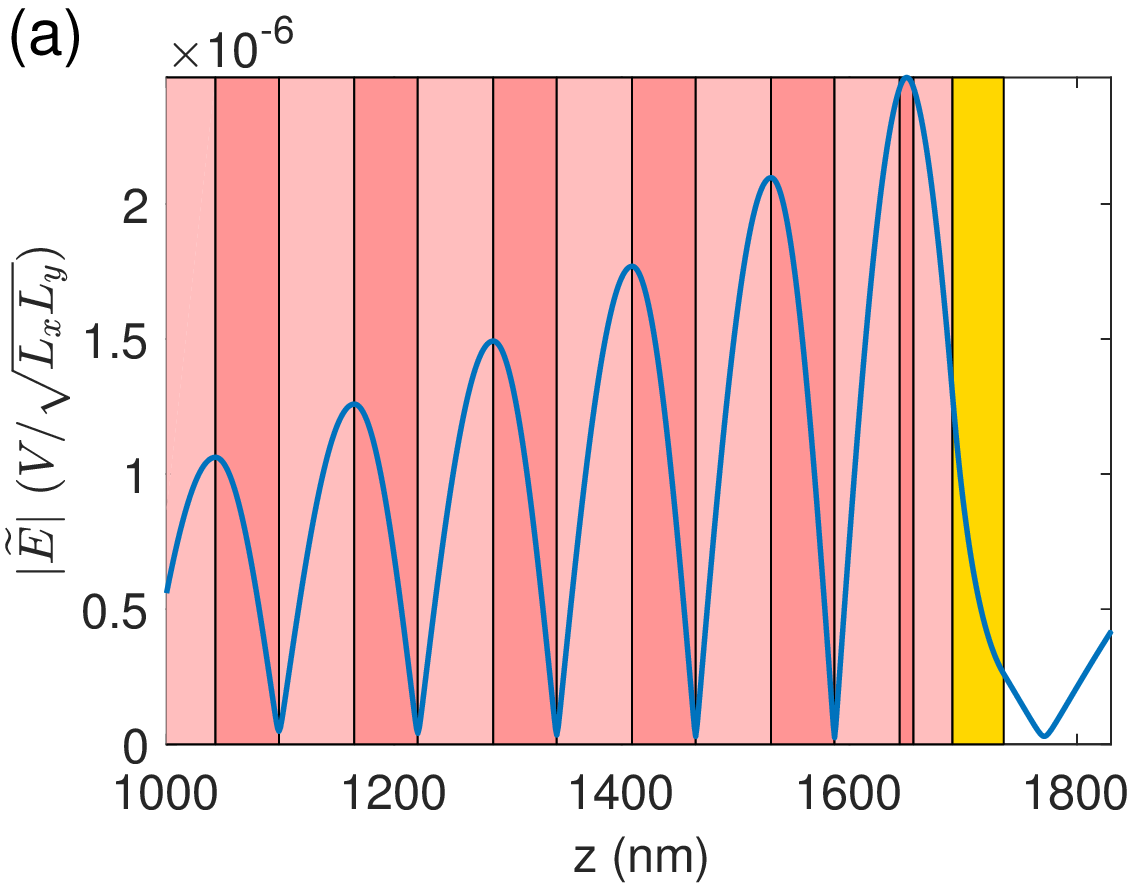}
	\includegraphics[width=0.235\textwidth]{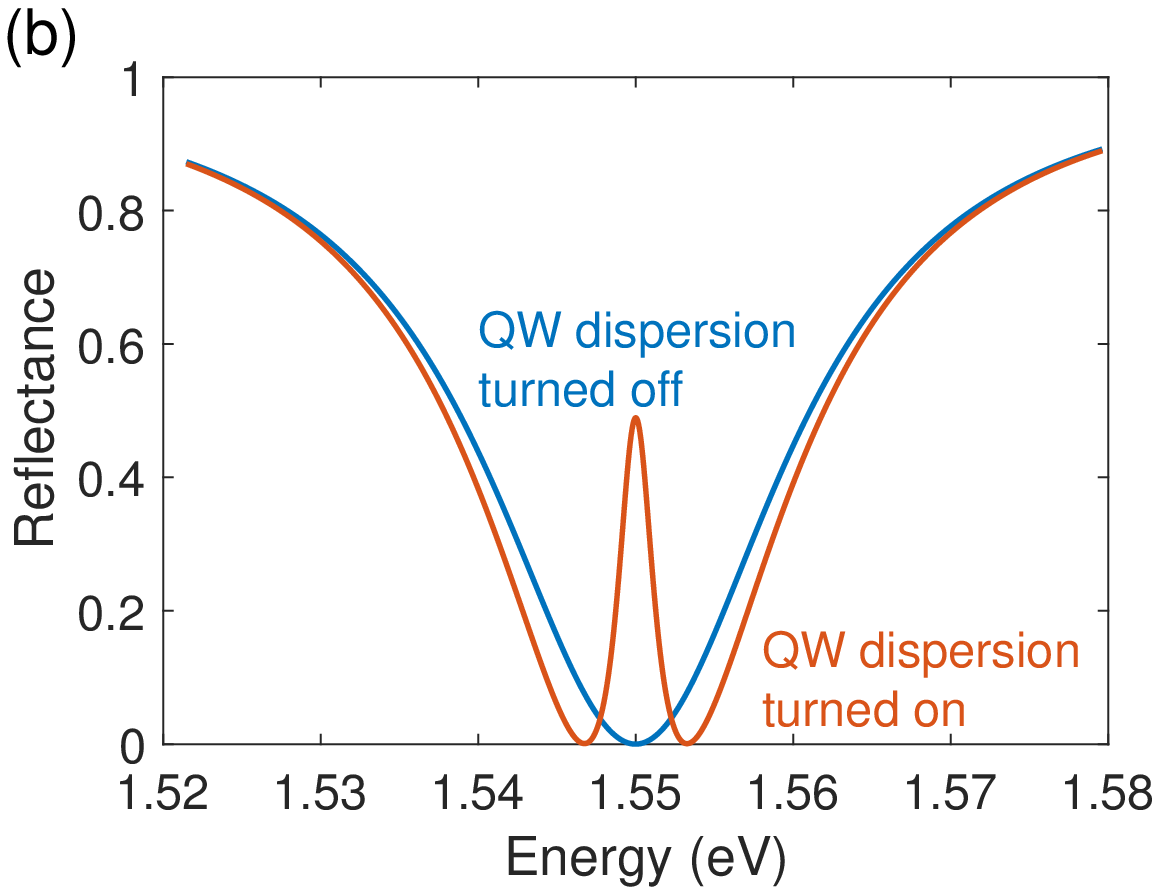}
	\caption{Vacuum field profile (a) and Rabi splitting (b) for Tamm-plasmon type cavity. The DBR structure is the same with previous DBR cavities. The excitation wave is launched from the substrate to top (left to right). The gold colored region represents a 45nm gold layer. Its refractive index is taken from Olmon, et al\cite{Olmon_Optical_2012} for evaporated gold at 800nm. The cavity Q is around 60 inferred from the linewidth, mainly due to the loss in the metal. The polariton splitting is 6.51meV, 43\% larger than conventional DBR cavity.
	}
	\label{fig:Tamm_DBR_cavity}
\end{figure}

To obtain even tighter field confinement, we consider another two types of cavities where the DBR is replaced by mirrors with shorter penetration depth -- a metal mirror and a sub-wavelength grating mirror. Metal mirror, with a typical penetration depth of less than 100~nm, can be used to form Tamm-plasmon polaritons\cite{Kaliteevski_Tamm_2007}. The optimal metal-DBR cavity we find consists of 45~nm thick gold layer as the top mirror (\fref{fig:Tamm_DBR_cavity}(a)), the same bottom DBR as in the conventional DBR-DBR cavity is used, and a $\lambda/2$ AlAs cavity layer slightly adjusted to compensate for the change of the reflection phase from $ \pi $.

In this structure, the field decays rapidly in the gold layer, resulting in a much shorter effective cavity length (\fref{fig:Tamm_DBR_cavity}(a)), therefore a much higher vacuum field strength. We obtain $E_{max}= 2.47\times 10^{-6} \rm V/\sqrt{L_xL_y} $ \footnote{For strong dispersive media like metal, the energy density in field normalization equation~\ref{eq:normalization} is modified by Eq.6.126 in Jackson, Classical Electrodynamics, 3rd edition, 2001} and $\hbar\Omega=6.51$~meV, 43\% larger than conventional DBR cavity. This is an impressive improvement considering only one side of the cavity is replaced by a metal mirror. The structure was used to create the first organic polariton \cite{Lidzey_Strong_2008, Lidzey_Room_2009}.  In III-As cavities, Grossmann et al\cite{Grossmann_Tuneable_2011} additionally replaced the bottom DBR with an air DBR to further improve the vacuum Rabi splitting to about double that of the conventional DBR cavity.\footnote{We have not been able to reproduce the field profile or reflection spectrum shown in Ref.~\cite{Grossmann_Tuneable_2011} in our simulation and instead found the 45~nm thick gold layer as the optimal for a Tamm-plasmon mirror.} However, the drawback of this structure is the low cavity Q due to metal loss. The cavity linewidth is comparable to the Rabi splitting; the system remains in the intermediate coupling regime rather than the strong coupling regime (\fref{fig:Tamm_DBR_cavity}(b))\cite{savona_quantum_1995}. The large loss in metal also makes it difficult to achieve quantum degeneracy or low threshold laser.

\begin{figure}[t]
	%\flushright % declaration
	\includegraphics[width=0.235\textwidth]{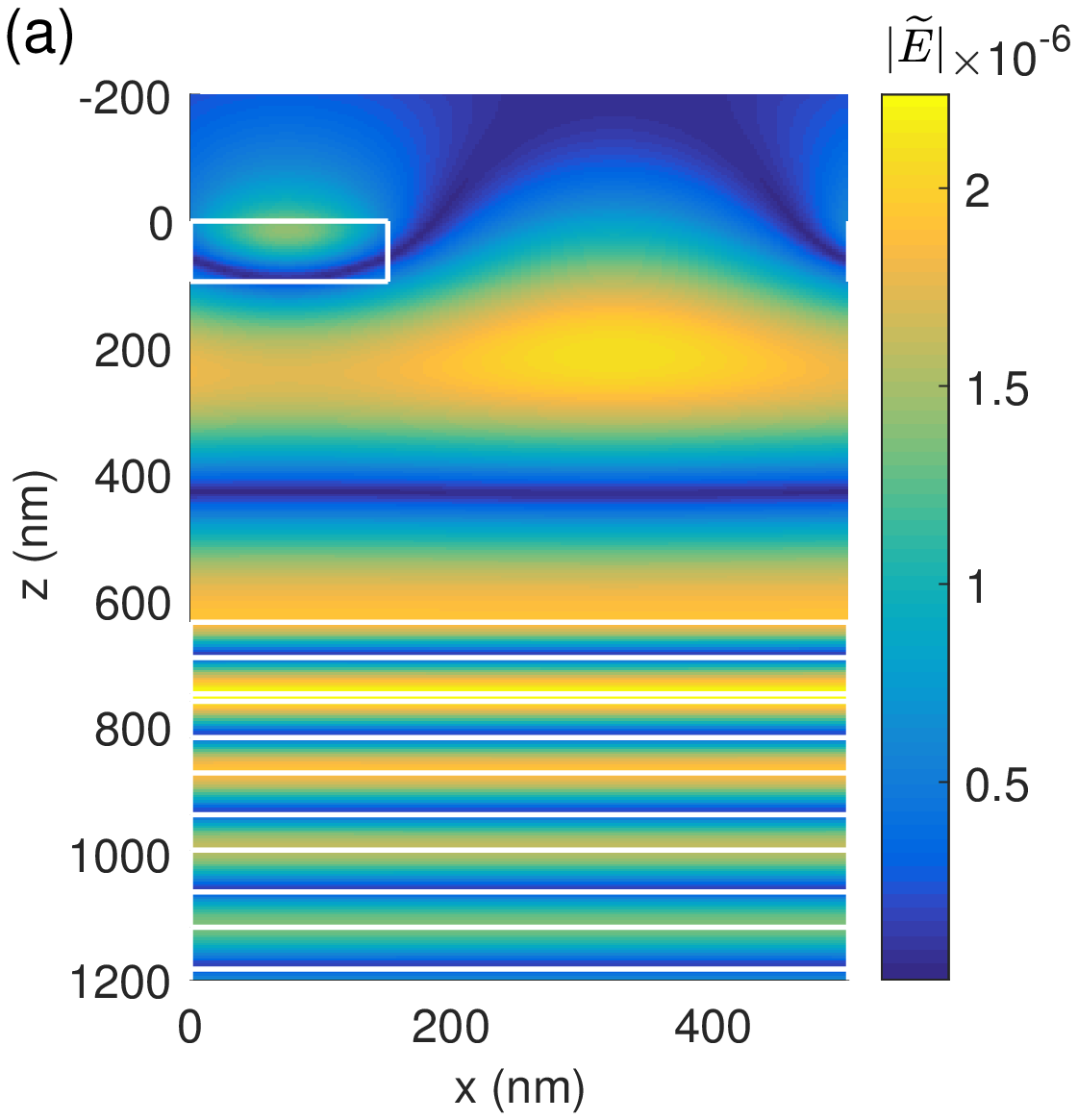}
	\includegraphics[width=0.235\textwidth]{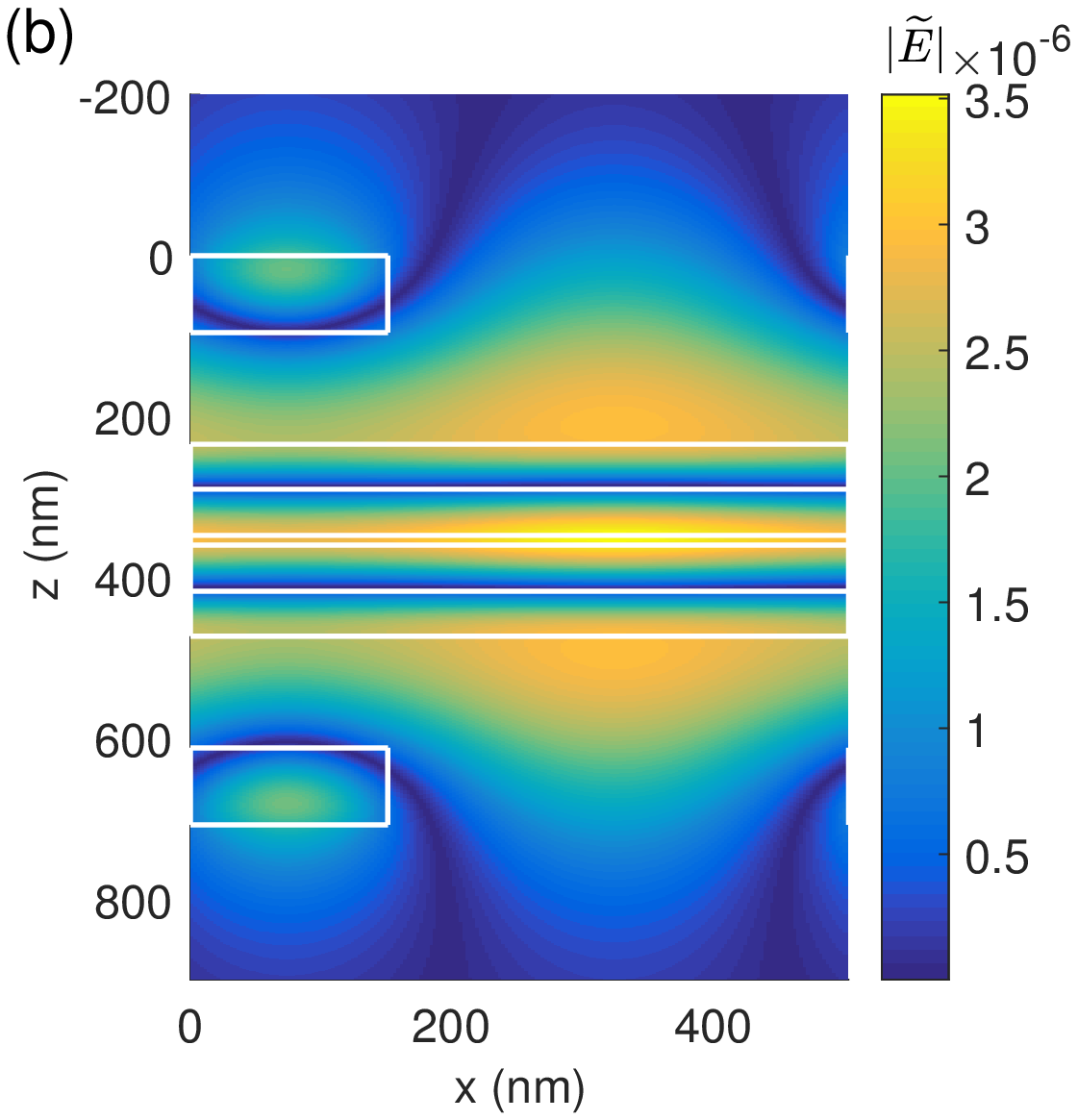}
	\includegraphics[width=0.235\textwidth]{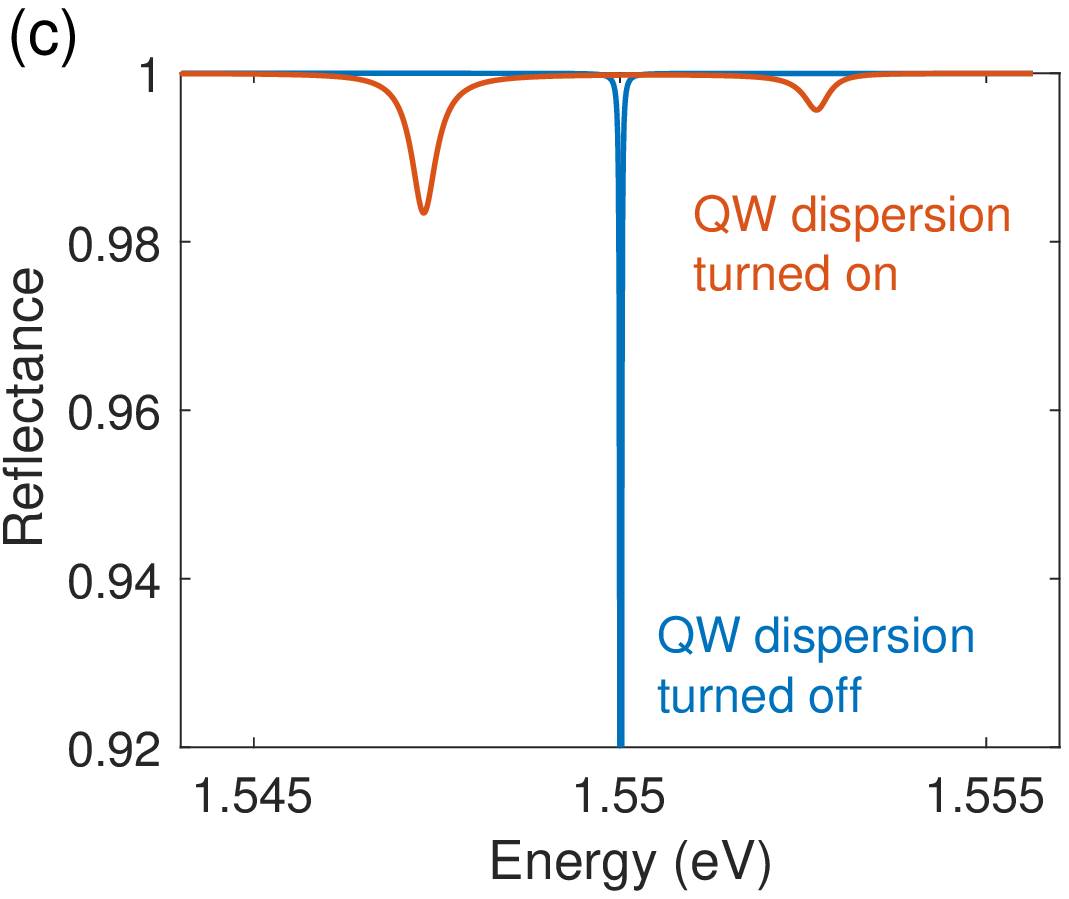}
	\includegraphics[width=0.235\textwidth]{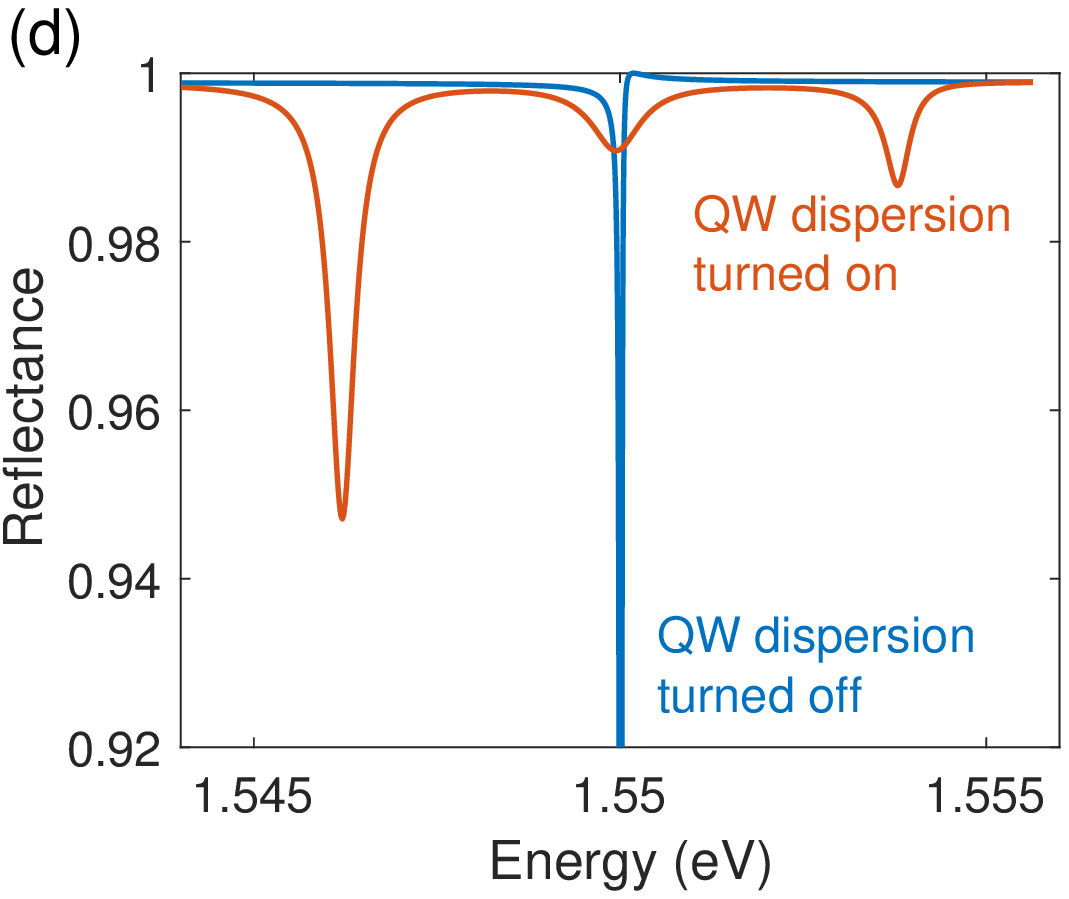}
	\caption{Vacuum field profile and Rabi splitting for SWG-DBR hybrid cavity (a-b) and double SWG cavity (c-d). The design parameters for the SWG are, thickness of grating 95.6nm, period 503nm, filling factor 0.3, using $ \rm Al_{0.15}GaAs  (n=3.58)$, with air gap 511nm (left) and 144nm (right) between SWG and the cavity layer. Bottom DBR in the hybrid cavity is the same with conventional DBR cavity used above. The middle resonance in the reflection spectrum (d) is due to the exciton absorption of the first-order grating diffraction mode (evanescent in air but propagating in the cavity layer).
	}
	\label{fig:SWG_cavity}
\end{figure}

To maintain both a high cavity $Q$ and short cavity length, we use a III-As subwavelength grating (SWG) for the fourth type of cavity \cite{huang_surface-emitting_2007, zhang_zero-dimensional_2014}. The SWG is a thin layer of high-index dielectric grating suspended in air, acting as the high-reflective mirror (\fref{fig:SWG_cavity}(a)).
To avoid non-zero order evanescent diffraction waves entering the cavity layer, we include a $ 3/4 \lambda $ air-gap between the SWG and the bottom structure. The $ \lambda/2 $ cavity layer and bottom DBR are identical to the conventional DBR cavity as in \fref{fig:air_DBR_cavity}(a). We obtain $E_{max}= 2.23\times 10^{-6} \rm V/\sqrt{L_xL_y} $ and $\hbar\Omega = 5.37$~meV, both 19\% larger compared to the conventional DBR cavity and comparable to the single-sided air-DBR cavity. The advantage of the SWG-DBR cavity is its high cavity quality and simplicity in fabrication. The structure is first fabricated as a high-quality, monolithic crystal by epitaxial growth, as with the conventional DBR cavities. The SWG is then created by lithography and etching. Because the grating is the only air-suspended layer created, SWG-DBR hybrid cavity is easier to make and more robust mechanically than an air-DBR. Polariton lasing and excellent coherence properties have been demonstrated in SWG-DBR cavities \cite{zhang_zero-dimensional_2014,fischer_magneto-exciton-polariton_2014,kim_coherent_2016}, but not yet in cavities using air-DBR or metal mirrors. 

Tighter cavity field confinement can be achieved when replacing the bottom DBR also by an SWG. As shown in \fref{fig:SWG_cavity} (b), a $ \lambda/2 $ AlAs cavity layer sandwiched by two $ \rm Al_{0.15}Ga_{0.85}As$ $ \lambda/4 $ layers is air-suspended between two identical SWGs.
Because of the evanescent diffraction waves from the SWGs, the field at the QW is slightly non-uniform. We record an average vacuum strength at the QW to be $ 2.92\times 10^{-6} \rm V/\sqrt{L_xL_y} $. The vacuum Rabi splitting is $\hbar\Omega = 7.58$~meV, 67\% larger than that of the conventional DBR cavity. 

The fabrication of such a double-SWG cavity is more challenging than the SWG-DBR hybrid cavity, but within reach of established fabrication techniques. The top-SWG and the rest of the structure can be fabricated separately at first, then they can be bonded together by cold welding \cite{Kena-Cohen_Strong_2008} or stacked together using the micromanipulation technique applied to making 3D photonic crystals\cite{Aoki_Coupling_2008}. Alternatively, one could make the SWGs from Si/SiO2 wafer, then do a wafer bonding with the III-V active layer\cite{Sciancalepore_Thermal_2012}. 
Additional practical challenges include heat dissipation and proper design of current injection path for electrical-pumped polariton devices.
Finally we note that the SWG-based cavity may be particularly useful for coupling to 2D materials, where similar procedures of integration are already necessary. Additionally, it is possible to hold 2D material with lower index materials, further enhancing the vacuum field.

%\section{Summary and Conclusion}
In summary, we have investigated the performance of strong coupling for several types of planar cavities. We use maximum vacuum field strength as the main figure of merit, in place of the effective cavity length or mode volume. The Tamm-DBR cavity can provide $ \sim $ 43\% improvement in Rabi splitting with only one mirror replaced by a gold thin film. Researchers using active media with a large oscillator strength may find this structure easy to fabricate and yet still sufficient for reaching strong-coupling. However, its low quality factor due to metal loss limits the coherence of the exciton-photon coupling and the resulting polariton modes. The air-DBR type cavities are shown to provide 21\% (55\%) improvement over conventional DBR cavities if one mirror (both mirror) is replaced with an air-DBR. SWG based cavities provide 19\% (67\%) improvement when one (two) SWG is used, comparable to (better than) that of air-DBR cavities. 
In practice, among all alternative structures to conventional DBR-DBR cavities, SWG-DBR cavities are so far the only ones where high cavity quality factor and polariton lasing have been demonstrated experimentally.
Our findings here may guide the design of cavity systems for the research and applications of semiconductor exciton-polaritons, especially for increasing their operating temperature, lowering the density threshold for quantum phase transition and polariton lasing, and incorporating newer materials.

ZW, HD acknowledge the support by the National Science Foundation (NSF) under Awards DMR 1150593 and the Air Force Office of Scientific Research under Awards FA9550-15-1-0240. We thank Pavel Kwiecien for his open-source RCWA code used for calculations in this work.

%\section*{References}
\bibliographystyle{aipnum4-1}
%\bibliography{references}
%

\end{document}